# Study of a Plane Free Jet Exhausting from a Channel by Vortex in Cell Method.

Mohamed Ali Knani[a,1], Henri-Claude Boisson[b,2,*],
André Giovannini[b,3,] and Taieb Lili[a,4]

[a] Laboratoire de Mécanique des Fluides, Faculté des Sciences de Tunis, Université de Tunis, Tunisia.
[b] Institut de Mécanique des Fluides de Toulouse, Toulouse, France.[*]

**SUMMARY**

A two-dimensional simulation of a plane jet exhausting from a channel has been performed using the vortex in cell algorithm in the Reynolds number range of 100 to 900. The vorticity is generated on the wall of the entrance channel whose length has been fixed in order to obtain a fully developed velocity profile at the entry of the jet. The transient behaviour of the velocity field starting from rest has been observed until reaching a quasi steady regime. The mean value of the velocity field is compared with the results of a finite volume computation on the same mesh. The velocity fluctuations obtained using this method are analysed. Their effect on the mean flow is estimated to be smaller than the viscous effect.

## 1. INTRODUCTION

A lot has been published about vortex methods for predicting two-dimensional flows and these methods are considered nowadays as classical [1]. Vortex methods are supposed to be quite convenient for treating shear flows because of their ability to concentrate a computation effort on the flow region where needed. Their main advantage is the use of a Lagrangian displacement of the vortex particles. The present work is developed in the context of the Random Vortex Method due to Chorin[2]. It is originally a grid-free method that solves the Navier Stokes equation in vorticity variables. This means that the vortices are transported by the velocity and a random motion is superposed to take account of the physical viscous diffusion. The velocity field, at each step in time, may be computed directly using Biot - Savart law for a given

---

vorticity field. Various cases have been fully tested against analytical solutions or experimental results. Most of the cases treated in the literature concern periodic boundary condition flows, wall flows, pipes channels or wakes. One can cite the results of Ghoniem and Gagnon[3] in the case of a backward facing step in a laminar regime, those of Gagnon et al.[4] in single as well as double symmetrical backward-facing step channel at high Reynolds number, and the more recent ones of Mortazavi and Giovannini downstream of a plate separator [5]. All these results have been tested upon experimental results and the question of a numerical false diffusion has been treated.

The present paper aims at studying the behaviour of a plane jet exhausting from a channel and freely flowing out of a semi - infinite domain. The vortex-in-cell (VIC) method is used for this purpose. It combines the Lagrangian displacement of point vortices with the Eulerian determination of stream function on a fixed grid. This algorithm was developed before by Christiansen [7], Gagnon and Huang[8], Savoie [9] and applied to the boundary layer by Pellerin and Giovannini[10]. Its main advantage is to reduce considerably the computing time compared to the use of direct computation of velocity fields by the Biot-Savart law. Thus, the number of point vortices needed in order to achieve a better accuracy can be increased for a fixed computational effort.

Hitherto, the case of jets has been mainly studied in the early development phase; that is, before the first vortices created in the initial shear layers leave the computational domain [6]. Indeed, the jets are very sensitive to the outflow boundary conditions and some perturbations may be reflected into the domain when structures exit. Generally, when passing through this frontier, one considers that the free outgoing of vortices produces negligible perturbations on the inner flow. This numerical experiment has been performed to certify that this assumption is still valid for long computation times, and to test the ability of the solver to obtain a basic time averaged flow sufficiently independent of the numerical perturbations. The purpose is to ensure valuable subsequent studies on instability and control.

The VIC computer code used by Savoie [9] has been extended to the case of a plane jet. The flow is analysed through the transient regime that leads to a steady state starting from the rest. The results are compared with those obtained using the finite volume (FV) method that was used by Knani et al. [11]. It is worthwhile noticing that we have also chosen to work in a regime ranging from moderate to low Reynolds numbers that corresponds to the unsteady laminar regime studied in the above mentioned paper [11], a situation in which the method adopted to treat the diffusion in this code has to be used with caution and the practical choice of parameters has to be carefully fixed. What is worth mentioning is that the comparison with a different numerical approach is well adapted to the fixed goal of the paper. The Eulerian finite volume Navier Stokes solver has been previously tested against experiments for the 2D plane jet [12] [13] Unfortunately, detailed experimental data on the near jet in the required conditions were not available. Nevertheless, elements of comparison with self similar solution are provided on the numerical results basis. In this paper, we present in succession the physical problem (section 2), the numerical method (section 3), the initial and boundary conditions (section 4), and the numerical results (section 5).

## 2. THE PHYSICAL PROBLEM

The domain of study is the plane jet exhausting from an entrance channel in a semi-infinite domain filled with the same fluid and bounded upstream by a wall. The flow is incompressible, viscous and two-dimensional. It corresponds to the conditions that have been found in the Varieras experiments [14]. In this last work [14] a varicose mode of instability was observed for a parabolic velocity profile at the exhaust and the problem exhibits an axial symmetry. Thus, half a domain is taken for the numerical simulation. The Reynolds number, based on the uniform inlet velocity $U_0=1$ and the height of the channel $H=1$, ranges from 100 to 900.

The domain and boundary conditions are represented in Figure 1. The lateral extent of the domain is the width D containing the vertical wall and the jet exhaust aperture $H/2$. The total length in the longitudinal direction is such that the jet domain has a constant length L equal to 30 and the channel length $l$ is adapted to the value of the Reynolds number.. The width of the jet D is fixed in order to let the point vortices flow out through the lateral face. For the higher Reynolds numbers ($Re \geq 300$), this condition is reached for a domain width of 3H but for $Re < 300$ it is necessary to extend this width to 6H. The boundary conditions indicated in this figure are discussed in section *4*.

## 3. THE NUMERICAL METHOD

In 2-D flows the mixed Eulerian-Lagrangian method uses the non-null component of the velocity curl, the vorticity $\omega$, as the transport variable. By taking the curl of the Navier-Stokes equations and relying the fact that the divergence of the velocity vector is zero, the Helmholtz dimensionless vorticity transport equation is obtained, which expresses the vorticity transport by convection and diffusion.

$$\vec{\omega} = \nabla \wedge \vec{V} \quad \text{and} \quad \omega = \|\vec{\omega}\| \qquad (1)$$

$$\frac{\partial \omega}{\partial t} + (\vec{V} \cdot \nabla)\omega = \frac{1}{Re}\Delta \omega \qquad (2)$$

where Re is the Reynolds number of the flow.

This last equation is coupled with the Poisson equation for the stream function :

$$\Delta \psi = \omega \qquad (3)$$

where $\psi$ is the stream function defined as :

$$\psi \Leftrightarrow \vec{V} = \nabla \wedge (\psi \vec{k}) \tag{4}$$

$\vec{k} \rightarrow$ unit vector of the normal to the jet plane.

To solve the equation (2) we use a technique proposed by Chorin [2] and widely studied by Beale and Majda [15]: the vorticity transport equation is solved by splitting it in into convection and diffusion operators. Both equations are treated separately in two fractional steps:

First step - Convective part:

$$\frac{\partial \omega}{\partial t} + (\vec{V} \cdot \nabla)\omega = 0 \tag{5}$$

Second step - Diffusive part:

$$\frac{\partial \omega}{\partial t} = \frac{1}{Re} \Delta \omega \tag{6}$$

The vorticity field is discretized into a finite number of vortices $N_v$. For a given time step, the total displacement is obtained by adding the convective $L(\vec{u}_i(x,y,t,\Delta t), dt)$ and the diffusive movements $\vec{\eta}$.

$$\vec{\chi}_i(x,y,t+dt) = \vec{\chi}_i(x,y,t) + L(\vec{u}_i(x,y,t,\Delta t), dt) + \vec{\eta} \tag{7}$$

$L(\vec{u}_i(x,y,t,\Delta t), dt)$ represents the integration operator for the velocity field.

The velocities of each point vortex are computed by a bilinear interpolation between the velocities over the grid. The diffusion step is treated by the Random Walk algorithm (Ghoniem and Gagnon [3] and Ghoniem[14] ). The diffusive transport $\vec{\eta}$ is then statistically simulated from a random process and a Gaussian distribution obtained from the two-dimensional Green solution of the equation (6). The independent random numbers $(\eta_x, \eta_y)$ are generated at each step in time with a fixed standard deviation $\sigma = \sqrt{2\Delta t / Re}$ [8]. The time marching algorithm adopted here corresponds to the first order Euler scheme.

The stream function is obtained by Vortex-In-Cell (VIC) algorithm proposed by Christiansen [7] which associates an Eulerian grid to the initial Lagrangian formulation. The vorticity resulting from the Lagrangian displacement is distributed on a fixed grid and serves to compute the vorticity by solving equation (3).

This latter equation is solved over this grid using a second order finite difference scheme and a SOR solver (successive-over-relaxation). The relaxation parameter is fixed, after optimisation, to 1.98 and convergence is achieved at $10^{-7}$. The grid size is fixed in order to obtain a suitable resolution in both directions. The criteria that were applied to select the appropriate number of points are grid independence, correct generation of vortices at the wall and outflow of the point vortices in the vertical outlet section. The selected mesh in such conditions is 840x60 for the higher Reynolds numbers (Re$\geq$ 300) and 840x120 for Re<300. In the first case, the height of the domain is D=3 and for the lowest values, the computation domain is D=6 for the same mesh size in the transverse direction.

Globally, the total computational time is proportional to the number of vortices $N_v$. This number is fixed by the reference circulation. Numerical parameters are kept constant in the simulation for all Reynolds numbers. The circulation reference is chosen to be $\Gamma_0=5.10^{-5}$. The method for the generation and the outflow of the vortices is described in section 5 consecrated to boundary conditions. This value of $\Gamma_0$ ensures a correct convergence of the method for a time step $\Delta t= .1$ according to Mortazavi et al. [17]. The linear evolution of the circulation and of the number of vortices in the transient phase before the vortices can leave the domain is in agreement with the theoretical growth of the vorticity at the wall and this is considered to be a confirmation of a correct behaviour of the numerical solver (see section 5).

## 4. INITIAL AND BOUNDARY CONDITIONS

The input velocity profile on BC_1 is unchanged throughout the computation. Starting with a potential flow, the vorticity is generated along solid walls in order to annihilate the slip velocity where no slip condition is prescribed (BC_2). The number of vortices introduced corresponds to a uniform distribution of the circulation on a wall cell length. The sign of elementary vortices of circulation $\Gamma_0$ depends on the local slip velocity and can be either positive or negative.

The boundary conditions are presented in Table 1 for both the VIC and the finite volume (FV) methods. The entry section of the jet is situated at x=0. For the VIC method, it is necessary to ensure that the repartition of point vortices in this section is not arbitrary and that it is compatible with the velocity profile. A channel of length $l$ is added at the entrance where a uniform longitudinal velocity profile is imposed. The vortices are generated from the wall of this channel and convected downstream. If the length $l$ is too short, an overshoot of longitudinal velocity is found at the entry of the jet. This disappears when the flow is fully established in the channel and the entrance section is well covered by point vortices. It is chosen to work in these conditions obtained for a length of at least $l$=12. This point is discussed further in section 5.

The upper boundary is the entrainment face, BC_4. It is necessary to let the flow enter freely. To this end a Neumann boundary condition for the normal velocity component is applied via the stream function and the vortices generated in the domain are cancelled if they accidentally reach this boundary.

The outflow boundary condition BC_5 is of major importance in the jet flow. We have adopted the policy to consider that the point vortices are simply cancelled when they overpass this boundary. One condition is needed: for the stream function the streamwise derivative of the normal velocity component is set to zero. But as the flow is perturbed near this boundary, the physically meaning part of the jet is restricted to an inner zone. A domain length L=30 is chosen but the analysis extends to a zone corresponding to $L_c \leq 25$.

The jet axis is treated as a symmetry axis BC_6. The stream function is fixed at a constant value on the axis of the channel and the vortices with negative ordinates are reflected (symmetric position and opposite vorticity sign).

In this paper we have compared the results obtained using VIC to those obtained by a finite volume method used by Knani et al. [11] for the same mesh and jet domain. The velocity profile is prescribed at the entry of the jet (x=0) and fixed as a parabolic profile. The other conditions are given in Table 1. It is found out with this numerical code that starting from rest at Re=500, a transient is observed, which vanishes after a given period.

## 5. THE NUMERICAL RESULTS

### 5.1. Evolution of vorticity and velocity fields

The number of vortices $N_v$ imposed the first step in time is fixed by the initial circulation and equal to $l/\Gamma_o = 2.4 \times 10^5$. These vortices are uniformly distributed on the wall of the channel. Then positive and negative vortices are generated in the course of the computation and convected by the velocity field. After a transient phase of rapid increase, both the number of vortices and the absolute value of the circulation reach a limit and start to oscillate (Figure 2.a and 2.b). Equilibrium is reached between the vortices shed by the wall and those leaving the domain. One observes that this equilibrium is reached at t=80 starting from the rest for Re=500.

At the Reynolds number range Re≤ 500, during the transient phase, the longitudinal velocity signals (Figure 3) show that a large perturbation is observed until t=50. It corresponds to a large transient structure that flows out rapidly. A quasi-steady regime seems to be reached afterwards. However, some small fluctuations are observed. Conversely for a Reynolds number of 900 or more, the steady regime is not observed and a fluctuating motion seems to persist resulting from the natural instability of the flow.

The point vortices positions in the domain are represented in Figure 4 for Reynolds numbers 100, 500, 900 at t=180 corresponding to the final time station of the computation. This figure illustrates the behaviour of the vortices in the domain (locations of 1/20 point vortices are represented). First, inside the channel, it can be seen that the establishment length is increasing with Reynolds number. In all cases the exhaust section is well fed with vortices and a parabolic velocity profile is expected. Second, the spreading of the jet as well as the unsteady pattern at Re=900 can be visualized by observing the vortex locations.

### 5.2. Velocity field in the jet

In Figure 5, the mean longitudinal velocity on the axis is displayed for different Reynolds numbers and over the whole length of the domain (from $-l$ to L). The mean values are calculated by time integration from t=100 to 180. The flow in the channel is fully established when the velocity on the axis reaches the limit of parabolic flow ($U_0$=1.5). For the value of $l$=20 displayed on the figure, this condition is satisfied for Reynolds numbers less than or equal to 500 and it is necessary to impose a length of 30 for higher Reynolds numbers. For Re=600 the results are compared with $l$=20 and $l$=30. The rate of the velocity decrease on the jet axis depends on the profile at the entry section (x=0). A slight difference exists in this case between the profiles due to the different lengths of the tube. This difference is greater for a Reynolds number of 900 (not shown on the figure).

The longitudinal velocities normalised with respect to the maximum initial value on the axis $(U_a)_0$ are plotted versus the transverse coordinate y in Figure 6 for Re=500 ($l$=12) and for different stations in the jet. The comparison of both methods shows that these mean velocity profiles are quite close to the instantaneous profiles of the finite volume method. Indeed, a quasi steady regime is obtained with the finite volume method (FV). As deduced from Figure 5, the results do not coincide for both methods and the profile at Re=500 for FV is nearer to the profile at Re=400 for VIC than the one at the same Reynolds number. This explains the differences noted on the value on the jet axis in Figure 6.

The results of figure 5 and 6 have been analyzed in terms of self similar solutions for laminar flow from the reference book of Schlichting [18]. Although the theoretical results retained for comparison are established for lower Reynolds numbers (Re<30), it is believed that the basic flow in the present case is closer to this regime than to the fully turbulent one that could have been considered otherwise. The conclusion is that self similarity is not completely reached in the domain and the velocity decrease law on the axis is gradually reaching the theoretical laminar law in $x^{-1/3}$ for Reynolds numbers up to 500 in the zone from 10H to 25H. Moreover, the velocity profiles collapse quite closely on a master curve $U/Ua = 1 - \tanh^2(\alpha y/y_{1/2})$ in which $y_{1/2}$ is such that $U/Ua = 1/2$. This behaviour is typical of the jet in self similarity and it has been found out by many authors even earlier than 1950 (see [18]). Similar results were found for plane jet with uniform velocity exhaust. Thus, although complete self similarity is not reached inside the domain, this result provides sufficient confidence in the correct evolution of the near zone of the jet.

It is also interesting to plot the mean vorticity map. The instantaneous vorticity is the algebraic sum of local point vortices in the domain at a given time station. This vorticity distributed over the grid points is used to compute the mean vorticity field over the time interval from 100 to 180. The iso-contours of $\omega_m$ (mean vorticity) over the external domain are represented in Figure 7. One can observe that this field is quite regular and does not exhibit large perturbations at the exhaust. The iso-lines are not distorted at the section x=20 which is considered as the end of the analysis domain in this work. The mean streamlines corresponding to the same averaging time interval are presented in figure 8. The entrainment flow rate on the top boundary $Q_0=(\psi_m)_0-(\psi_m)_{20}$ is equal to 0.225 for both methods. In this expression $(\psi_m)_0$ and $(\psi_m)_{20}$ are respectively the values of the mean stream function on this boundary for x=0 and x=20.

The fluctuations due to the vortices displacement induce local fluctuations of velocity that are observed in the computations. They are displayed in Figure 9, which represents the rms values of both u and v fluctuating velocity components, for x=10. In this section a maximum of $u'=\sqrt{\overline{u^2}}$ is found on the axis and it represents less than 4% of the initial input velocity. Another maximum is located near y=1; that is approximately the location of the shear layer issued from the jet lips. The maximum of $v'=\sqrt{\overline{v^2}}$ is situated in between and corresponds to a minimum of u'. This structure reproduces the situation that is observed inside the input channel in which two maxima of u' are observed near the wall and the axis. The value of v' is null at both of these locations as imposed by the boundary conditions. The non linear effect of such fluctuations produces a non-null but small $\sigma_{uv} = \overline{uv}$ correlation similar to a Reynolds stress effect. Figure 10 represents the lateral evolution of $\sigma_{uv}$ for different longitudinal stations. The absolute values of the curves negative peaks are smaller than $7 \times 10^{-5}$ all over the jet analysis domain. Of course the physical meaning of these fluctuations is not linked to 3D turbulence. However, in the near field of a plane jet with uniform velocity entrance exhibiting coherent vortices, the organised fluctuations have been found out to behave in a similar way in physical experiments [19], [20]. The effect of these fluctuations does not modify deeply the mean jet behaviour that remains similar to the continuous viscous flow obtained by the finite volume method. Indeed, within the analysis zone, the corresponding effective viscosity calculated by the ratio of the $\overline{uv}$ correlation and the longitudinal velocity gradient is less than $10^{-5}$, a value far smaller than the prescribed viscosity.

## 6. CONCLUSION

The numerical code based on Vortex-In-Cell algorithm used by [10] that has previously provided good results for boundary layers was adapted to the case of a plane jet in the Reynolds number range [100 - 900]. The mean velocity fields are in a reasonable agreement with an Eulerian finite volume approach [11]. They equally show good trends of a classical self similar behaviour for laminar jet [18].

In this study, we have adopted very simple conditions for the outflow and inflow. For this reason, it was necessary to check that they correspond to the one used in a similar Eulerian simulation. For the inflow, a variable length channel has been added to impose a controlled entrance condition in the domain. The generation of vortices comes from the wall of this channel. For the outflow, the vortices transported outside the computation domain are removed, which induces a finite perturbation in the exhaust section.

Our work has shown that the quasi steady state is reached for the mean value evaluated after the crossing of the initial transient at a Re=500. However, fluctuations are observed and are found out to vary in the lateral direction. Maximal intensities of longitudinal fluctuations are obtained in the jet shear layer and near the jet axis. These are partly linked to the random walk algorithm that is adopted for the diffusion step and partly to the jet instability. In the analysed Reynolds number range, marked by a high sensitivity to viscous effects, both methods provide similar jet behaviours. The simulation is considered as physically significant for periods corresponding to the crossing of several large vortices through the jet.

AKNOWLEDGEMENTS

The authors wish to thank the CNRS, the French Ministry of Foreign Affairs and the University of Tunis for financial support to this collaboration between the laboratories of both Countries.

# REFERENCES

1. Gustafson K.E. & Sethian J.A., Vortex methods and vortex motion, *SIAM, Philadelphia, Pennsylvania*, 1991.
2. Chorin A.J. .Numerical study of slightly viscous flow. *J. F. Mech.,* **57** , 1973, 785-796.
3. Ghoniem A.F. and Gagnon Y. Vortex simulation of laminar recirculating flows., *J. Comp. Phys.,* **68**, (2), 1987, 346-377.
4. Gagnon Y., Giovannini A. and Hébrard P., Numerical simulation and physical analysis of high Reynolds number recirculating flows behind sudden expansions., *Phys. Fluids A*, vol **5**, part 10, (1993), 2377-2389.
5. Mortazavi I. and Giovannini A., The simulation of vortex dynamics downstream of a plate separator using a vortex-finite element method., *Int. J. Fluid Dynamics*, 2001, vol.**5**, article 4, 41-58.
6. Kyia M., Ido, Y. & Akiyama H., Vortical structure in forced unsteady circular jet : Simulation by 3D vortex method., *ESAIM:Proceedings*, Vol **1**., 1996, 503-520.
7. Christiansen J.P. Numerical simulation of hydrodynamics per the method of point vortices. *J. Comp. Phys.*, **13** ,1973, 363-379.
8. Gagnon Y., Huang W., Fast vortex method for the simulation of flows inside channels with and without injection, *J. Thermal Science*, Vol. **2**, Nr. 1, 1993, 1-11.
9. Savoie R., Etude numérique d'écoulements de fluides par une méthode vortex : la marche descendante et les cavités sur les ailes de papillons., *PH. D thesis, Sherbrooke*, Canada, 1996.
10. Pellerin S. and Giovannini A. Interaction Vortex-Boundary layer : Numerical study of wall mechanism. *ESAIM Proceeding* , vol.**7** ,1999,325-334.
11. Knani M.A, Lili T. and Boisson H.C., Response of plane viscous jet to entrance flow rate perturbation., *Int. J. Numer. Meth. Fluids .*, 2001 , **37** , 361-374.
12. Meyer J., Sévrain A., Boisson H.C. and Ha Minh H., "Numerical simulation of a plane jet." Vieweg Verlag, 1990, *Notes on Numerical Fluid Mechanics*, Vol. 29, pp. 363-371, edited by P. Wesseling, Vieweg Verlag, Branschweigh, 1990.
13. Meyer J., Sévrain A., Boisson H.C. and Ha Minh H., "Organized structures and transition in the near field of a plane jet.", in *Structure of Turbulence and Drag Reduction*, édité par A. GYR, Springer Verlag, 1990
14. Varieras D., Etude de l'écoulement et du transfert de chaleur en situation d'impact de jet plan confiné, *Thèse Université Paul Sabatier*, 2000
15. Beale J.T. and Majda A. Rates of convergence for viscous splitting for the Navier-Stokes equations , *Math. Comp.*, **37**, 1981, 243-259.
16. Ghoniem A.F., Computational methods in turbulent reacting flow, *17th AMS/SIAM Summer Seminar on Reacting Flows* , Cornell University, Ithaca, 1985, 1-67.

17. Mortazavi I., Micheau P. and Giovannini A., Etude de la convergence numérique d'une méthode vortex pour un écoulement à grand nombre de Reynolds dans un mélangeur, *Comptes Rendus Mécanique, 2002, 330* , 6, 409-416.
18. Schlichting, H., Boundary layer theory, 7[th] edition,, Mac Graw Hill Book Company, New York (1979).
19. Faghani D., Etude des structures tourbillonnaires de la zone tourbillonnaire d'un jet plan : approche stationnaire multidimensionnelle., *Thèse INP Toulouse*, 2000.
20. Faghani D, A. Sévrain A., Boisson H.C., "Physical Eddy Recovery through Bi-Orthogonal Decomposition in an Acoustically Forced Plane Jet.", Flow, Turbulence and Combustion, 62**,** 69-88, 1999.

# APPENDIX A . NOMENCLATURE

| | |
|---|---|
| D | width of the domain |
| H | width of the jet or channel (=1) |
| $\vec{k}$ | unit vector of the normal to the jet plane |
| $L_0$ | length of whole computational domain |
| L | length of the jet domain |
| $L_c$ | length of the analysis domain |
| $l$ | length of channel |
| Nv | total number of point vortices |
| $Nv_+$ | number of positive point vortices |
| $Nv_-$ | number of negative point vortices |
| Re | Reynolds number ($U_0 H / \nu$) |
| t | current time |
| U | longitudinal component of velocity |
| $U_a$ | axis longitudinal velocity |
| $U_0$ | Reference velocity (=1) |
| $(U_a)_0$ | axis longitudinal velocity at x=0 |
| $U_m$ | mean longitudinal velocity |
| $u$ | longitudinal velocity fluctuation |
| $\vec{V}$ | velocity field |
| V | lateral component of velocity |
| $v$ | lateral velocity fluctuation |
| $u', v'$ | rms values of the fluctuating velocity components |
| x,y | space co-ordinates |
| $\overline{(a)}$ | Mean value of $a$ |

*Greek letters :*

| | |
|---|---|
| $\Gamma$ | circulation |
| $\Gamma_0$ | reference circulation |
| $\Gamma_+$ | positive circulation |
| $\Gamma_-$ | negative circulation |
| $\nu$ | kinematic viscosity (=1/Re) |
| $\omega$ | vorticity |
| $\omega_m$ | mean vorticity |
| $\psi$ | stream function |
| $\psi_m$ | mean stream function |
| $\chi$ | Lagrangian convective displacement |
| $\vec{\eta} = (\eta_x, \eta_y)$ | Lagrangian diffusive displacement |
| $\sigma$ | standard deviation of $\vec{\eta}$ |
| $\sigma_{uv}$ | cross correlation between velocity components $u$ and $v$ |
| $\Delta x, \Delta y, \Delta t$ | Eulerian mesh size and time step |

# List of Tables and Figures

## 1-Tables

Table 1 Boundary conditions

## 2- Figures

Figure 1 - Domain and boundary conditions

Figure 2 -Time Traces of: (a) Circulation; (b) Number of Vortices in the flow domain at Re=500 ($l/\Gamma_0$=2.4 $10^5$)

Figure 3 - Time traces of longitudinal velocity U, (Re=500 ; y=0.7)

Figure 4 - Instantaneous point vortices locations, (t=180)

Figure 5 - Longitudinal velocity on the x-axis, (t=180)

Figure 6- Longitudinal normalised mean velocity profiles
Re=500; $l$=12; $t \in [100, 180]$
*FV: instantaneous velocity at t=180.

Figure 7– Mean iso-vorticity contours
(Re=500; $t \in [100, 180]$ )
*(30 iso-contours equally spaced between 0 (axis) and –5.26)*

Figure 8– Mean iso-streamline function contours
(Re=500; $t \in [100, 180]$ )
*(22 iso-lines equally spaced between –0.25 (axis) and 0.3 and 8 iso-lines between 0.3 and 0.5 (upstream wall))*

Figure 9 – RMS of velocity fluctuations
(Re=500; x=10; $t \in [100, 180]$ )

Figure 10 – Cross correlation $\sigma_{uv}$ between velocity components
(Re=500; $t \in [100, 180]$ )

| Boundary | Vortex-in Cells | Finite Volume |
|---|---|---|
| 1 | $U(y) = U_0 = 1; V(y) = 0$ | *Not relevant* |
| 2 | $V(x) = 0; \psi(x) = 0$<br>*No slip → vortex generation* | *Not relevant* |
| 3 | $y \leq \dfrac{H}{2} \to U(y) \to$ *fluid domain*<br>$y > \dfrac{H}{2} \to U(y) = 0; \psi(y) = 0$ | $y \leq \dfrac{H}{2} \to U(y) =$ *parabolic*$; V(y) = 0$<br>$y > \dfrac{H}{2} \to U(y) = 0; V(y) = 0$ |
| 4 | $y = D \to \dfrac{\partial^2 \psi}{\partial x \partial y} = 0$<br>$\omega_i \to$ *free output* | $y = D \to U = 0; \dfrac{\partial V}{\partial y} = 0$ |
| 5 | $x = L \to \dfrac{\partial^2 \psi}{\partial x^2} = 0$<br>$\omega_i \to$ *free output* | $x = L \to \dfrac{\partial^2 U}{\partial x^2} = 0; \dfrac{\partial V}{\partial x} = 0$ |
| 6 | $y = 0$ & $-l < x < L \to$<br>$\psi = \psi(-l, 0)$<br>$\omega_i \to$ *mirror condition* | $y = 0$ & $0 < x < L \to V = 0; \dfrac{\partial U}{\partial y} = 0$ |

Table 1 - Boundary conditions

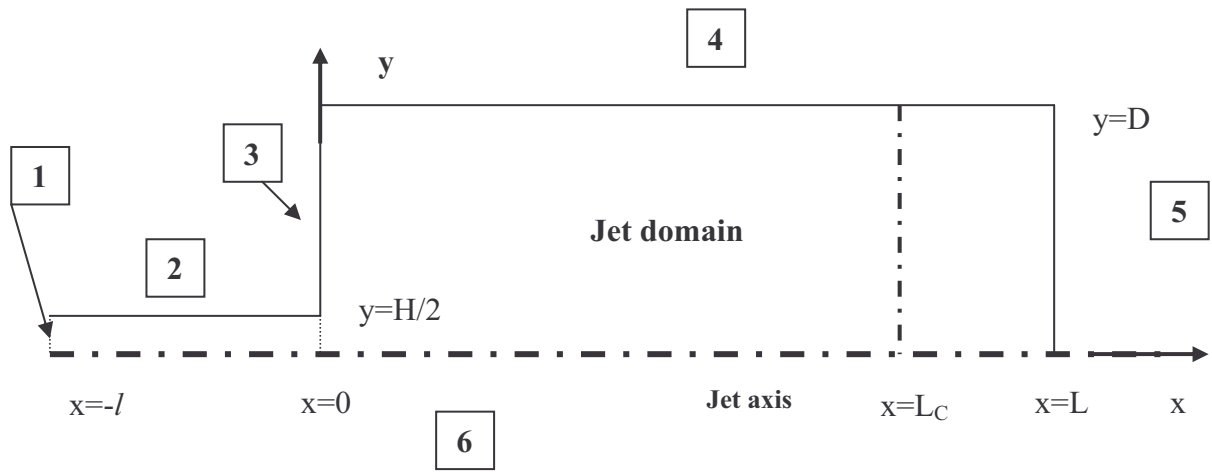

Figure 1- Domain and boundary conditions

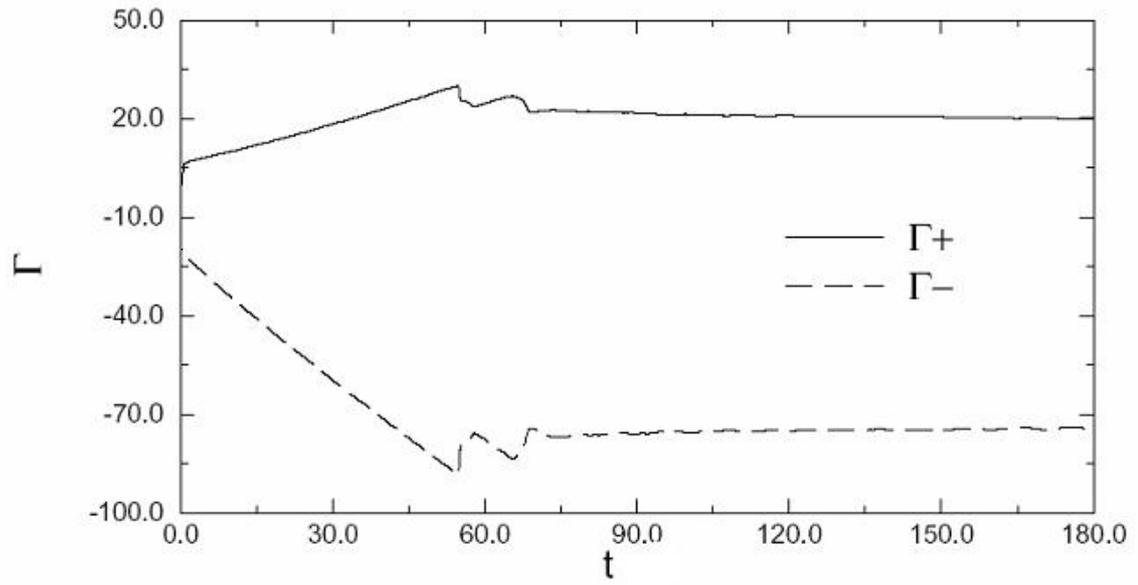

(a)

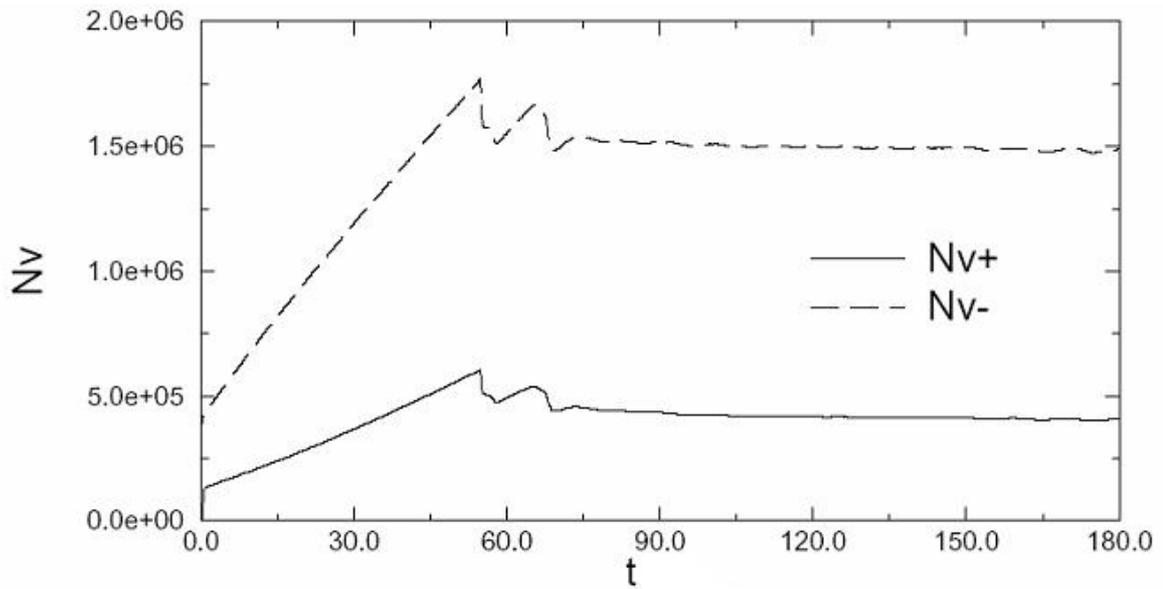

(b)

Figure 2 - Time Traces of: (a) Circulation; (b) Number of Vortices in the flow domain at Re=500 ($l/_{\Gamma\,0}$=2.5 $10^{-4}$)

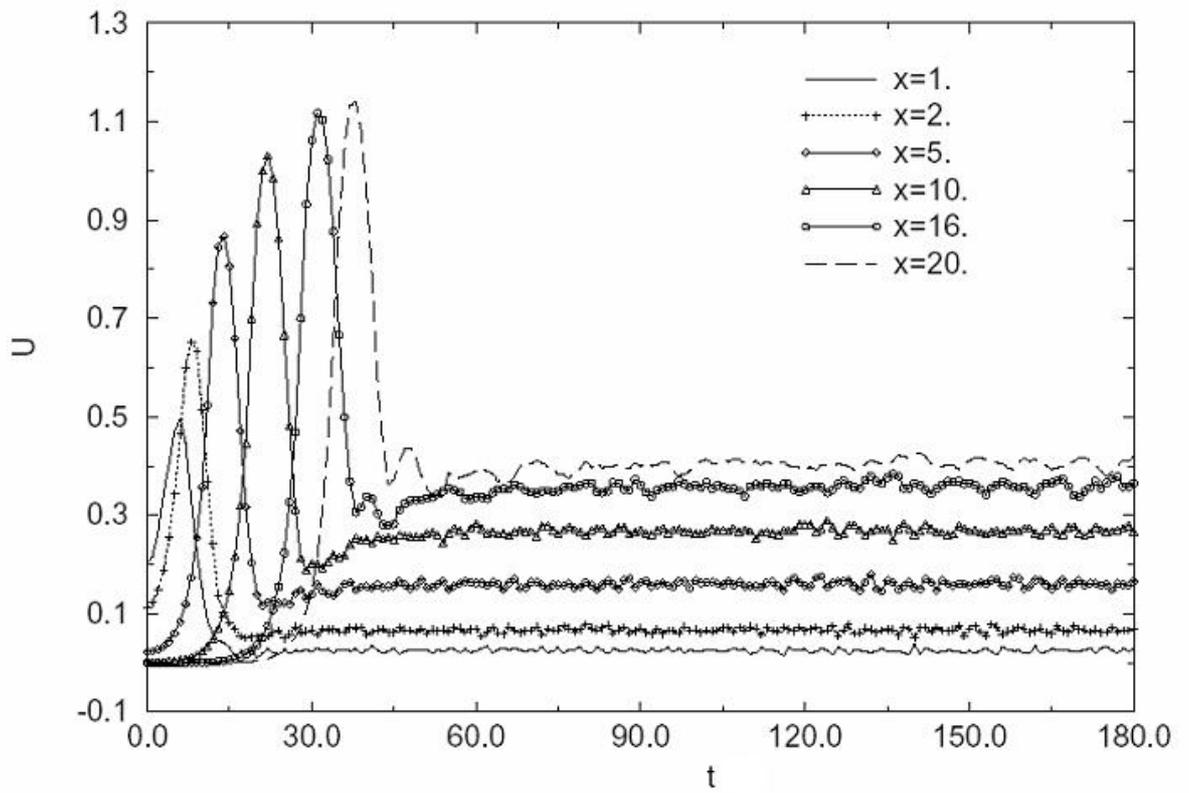

Figure 3 - Time traces of longitudinal velocity U
(Re=500 ; y=0.7)

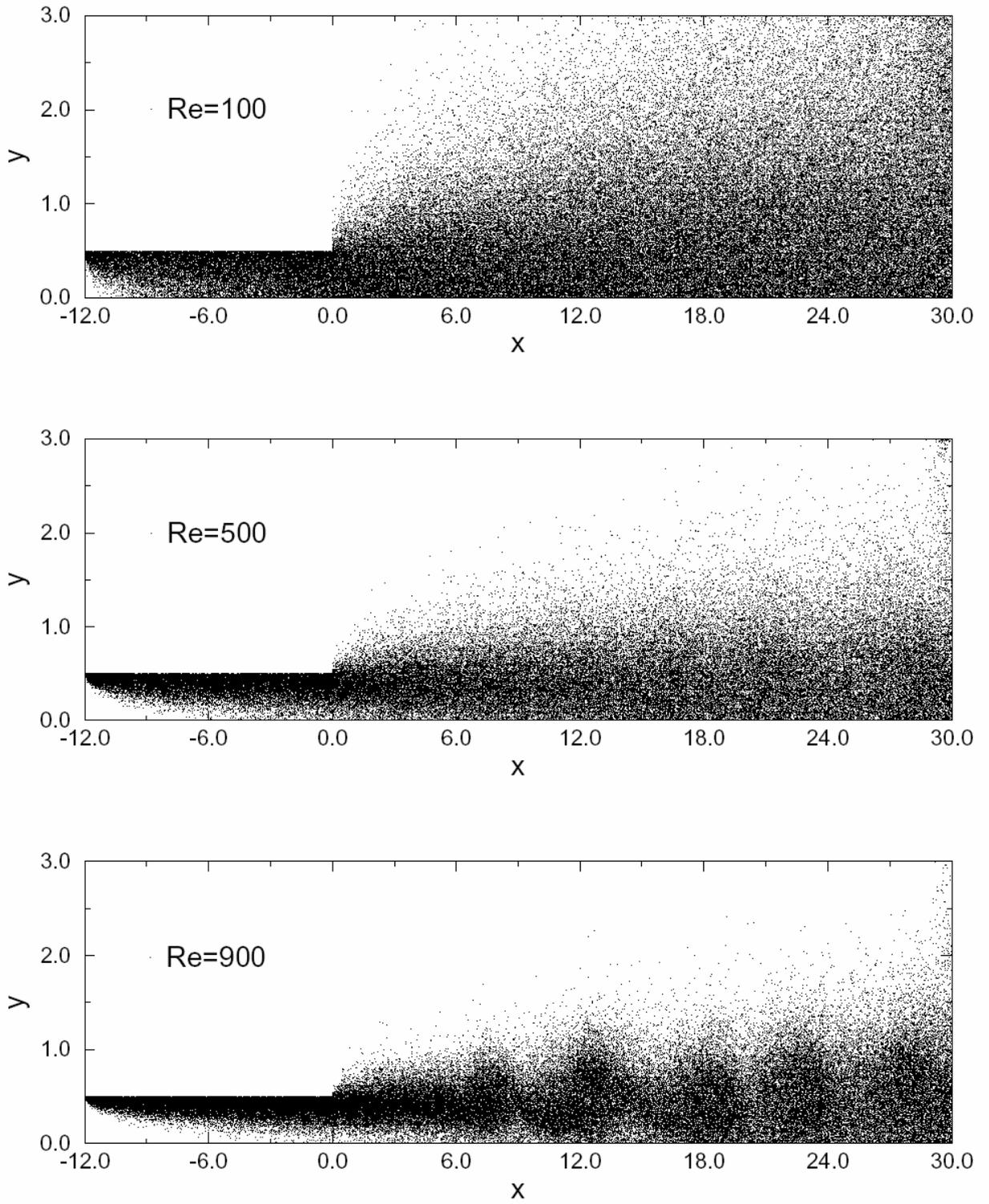

Figure 4 - Instantaneous point vortices locations

(t=180)

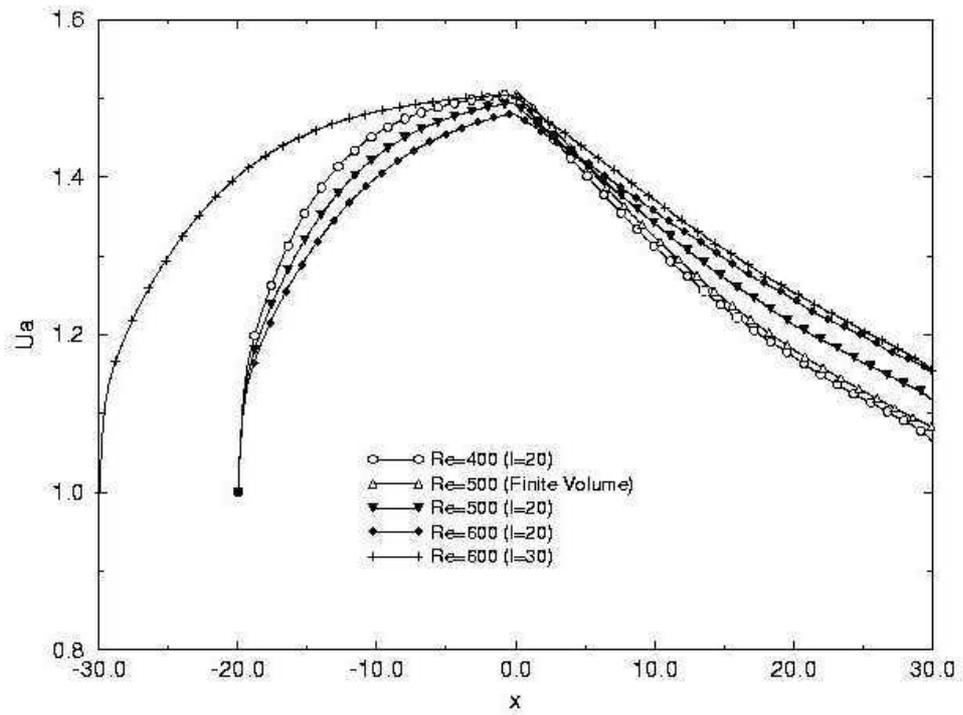

Figure 5 – Longitudinal mean velocity on the x-axis
(for t∈ [100,180])

25/29

Figure 6- Longitudinal normalised mean velocity profiles
Re=500 ; *l*=12; t$_\in$ [ 100, 180]
*FV : instantaneous velocity at t=180 .

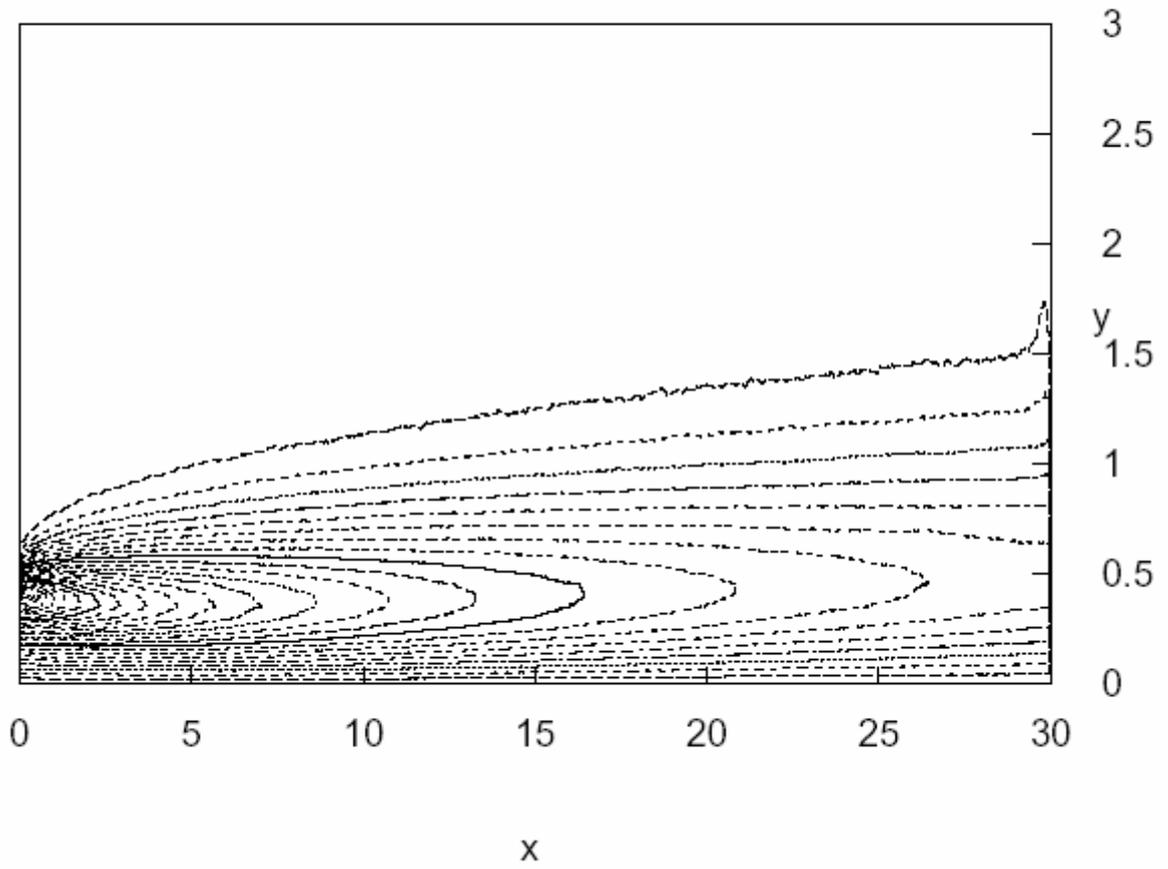

Figure 7– Mean iso-vorticity contours
(Re=500; t ∈ [100, 180])

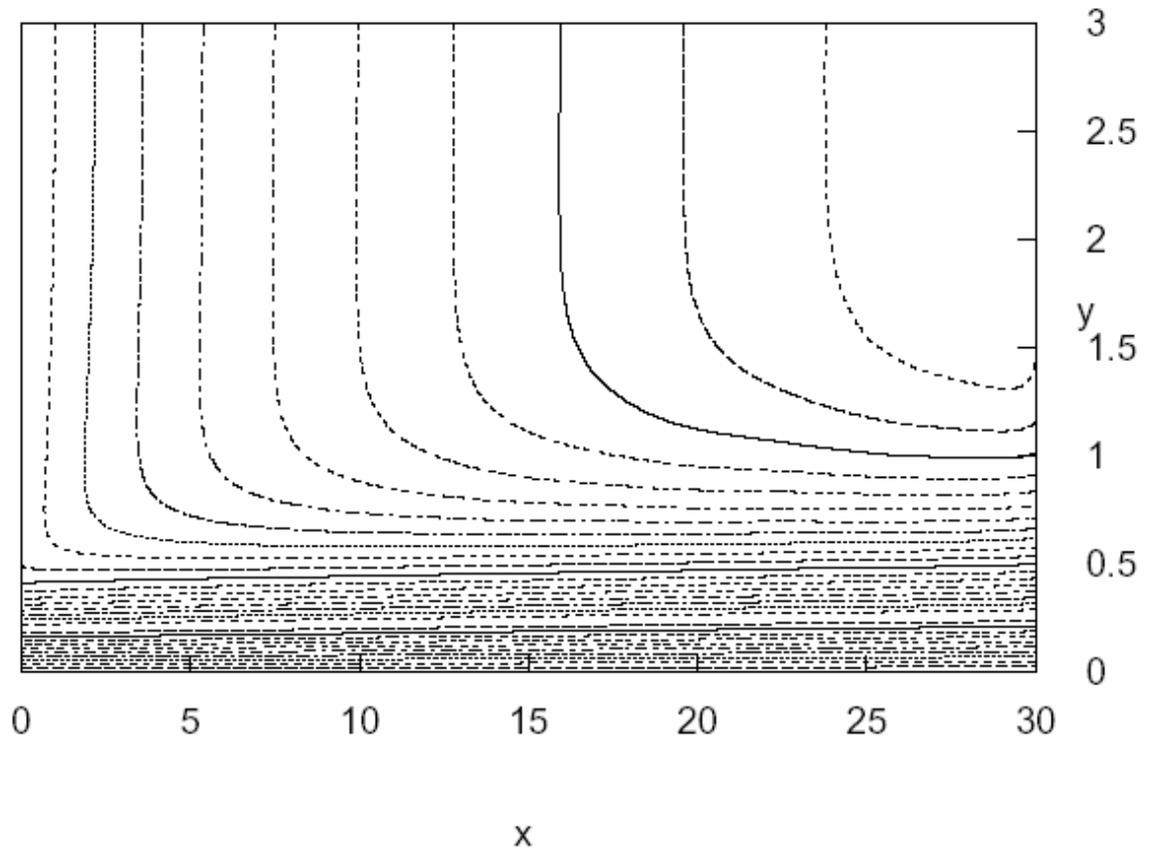

Figure 8– Mean iso-streamline function contours
(Re=500 ; t∈ [ 100, 180] )

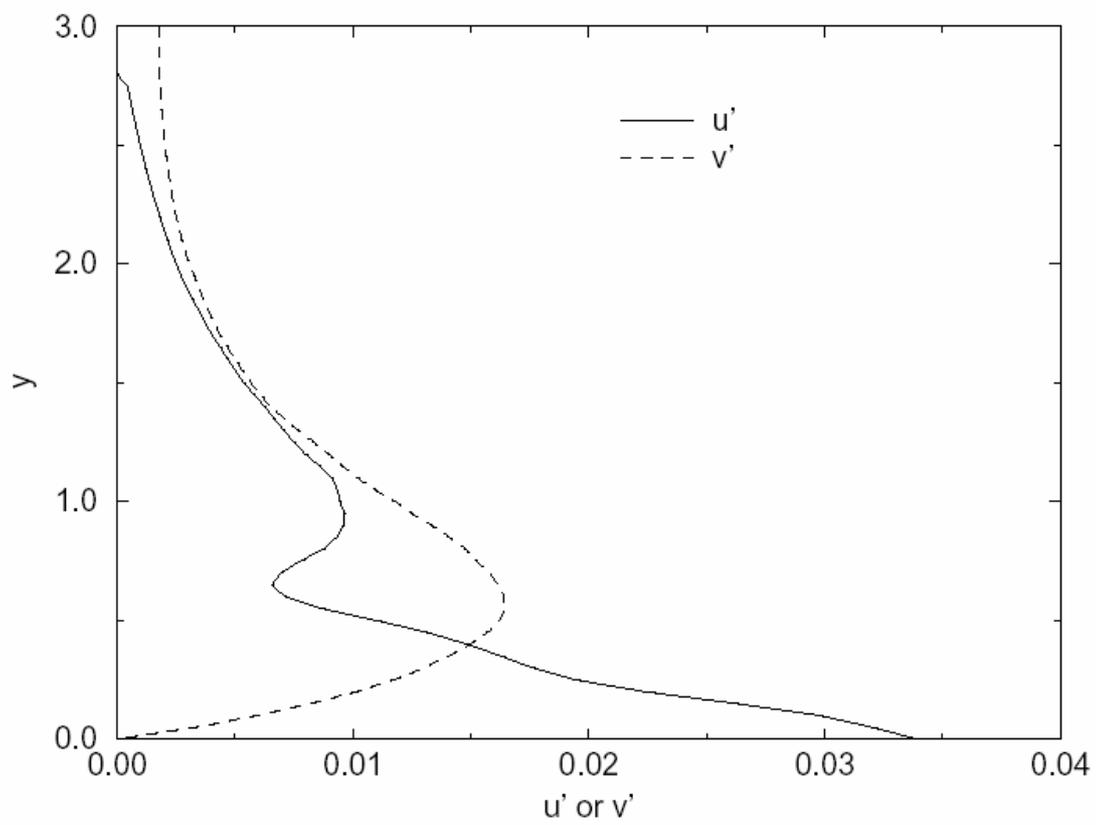

Figure 9 – RMS of velocity fluctuations
(Re=500; x=10; $t \in [100, 180]$ )

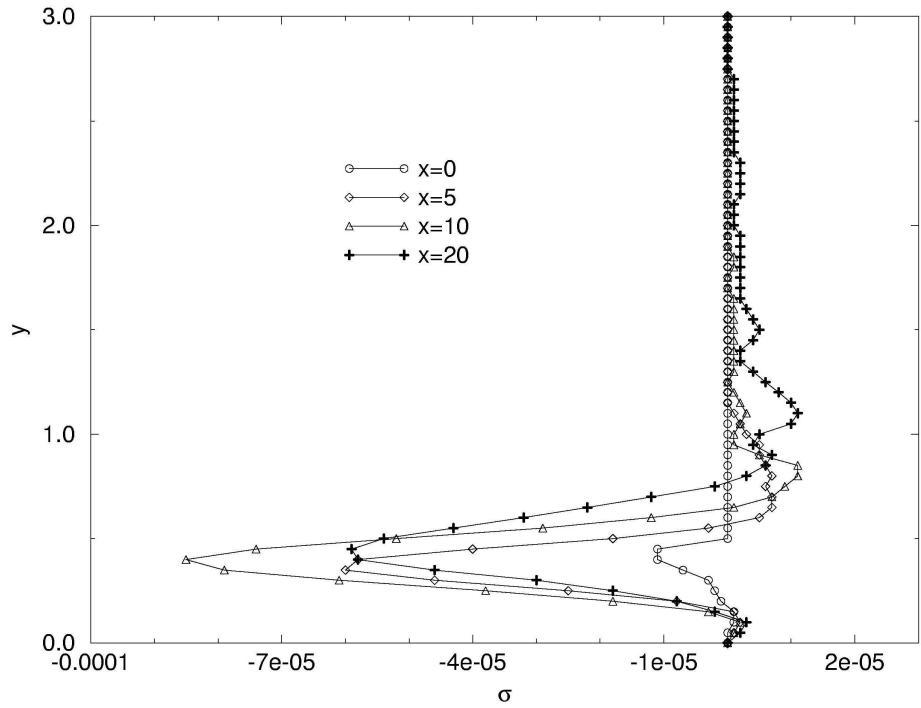

Fig. 10- Cross correlation $\sigma_{uv}$ between velocity components
(Re=500; $t \in [100, 180]$